\documentclass[prl,reprint,twocolumn,10pt,showpacs,superscriptaddress,nofootinbib]{revtex4-1}
\usepackage{amsmath}
\usepackage{amsthm}
\usepackage{graphicx}
\usepackage{times}
\usepackage{amssymb}
\usepackage{hyperref}
\hypersetup{
colorlinks=true,
linkcolor=red,
citecolor=red,}
\usepackage{textcomp}
\usepackage{xcolor}
\usepackage{enumerate}
\usepackage[final]{pdfpages}
\usepackage{multirow}
\usepackage{tabularx}
\usepackage{dcolumn}
\usepackage[mathscr]{euscript}
\usepackage{mathtools}
\usepackage{needspace}
\usepackage[mathscr]{euscript}
\usepackage{textcomp}
\usepackage{calrsfs}

\newcommand{\dmua}[1]{\int\frac{d^{2}#1}{\pi}}
\newcommand{\dmub}[2]{\int \frac{{d^{2}#1}{d^{2}#2}}{\pi^2}}

\begin{document}
\title{Quantum Correlations in Nonlocal BosonSampling}
\author{Farid Shahandeh}
\email{Electronic address: f.shahandeh@uq.edu.au}
\author{Austin P. Lund}
\author{Timothy C. Ralph}
 \affiliation{Centre for Quantum Computation and Communication Technology, School of Mathematics and Physics, University of Queensland, St Lucia, Queensland 4072, Australia}
\date{\today} 

\begin{abstract}
Determination of the quantum nature of correlations between two spatially separated systems plays a crucial role in quantum information science. 
Of particular interest is the questions of if and how these correlations enable quantum information protocols to be more powerful. 
Here, we report on a distributed quantum computation protocol in which the input and output quantum states are
considered to be classically correlated in quantum informatics.
Nevertheless, we show that the correlations between the outcomes of the measurements on the output state cannot be efficiently simulated using classical algorithms. 
Crucially, at the same time, local measurement outcomes can be efficiently simulated on classical computers. We show that the only known classicality criterion violated by the input and output states in our protocol is the one used in quantum optics, namely, phase-space nonclassicality.  
As a result, we argue that the global phase-space nonclassicality inherent within the output state of our protocol represents true quantum correlations.
\end{abstract}


\maketitle


\paragraph*{Introduction.---}

Correlations play an undeniable role in our understanding of the physical world.
In our macroscopic classical description of commonplace phenomena, classical physics and classical information theory are in perfect agreement in characterization and quantification of correlated events.
At a microscopic level, where quantum theory is our best candidate for explaining phenomena, however, the situation is different.
Our informational inspections of a quantum world, i.e., quantum information theory, is based on our intuition from classical information theory and classical probability theory, mostly the notion of quantum entropy~\cite{Nielsen,Misra15}.
However, a discrepancy emerges when physical constraints are taken into account to distinguish between quantum physics and classical physics, hence splitting quantum information approach to correlations from that of quantum optics.

In quantum optics it is common to study nonclassical features of bosonic systems in a quantum analogue of the classical phase space~\cite{Wigner}.
While in a classical statistical theory in phase-space the state of the system is represented by a probability distribution, the quantum phase-space distributions can have negative regions, and hence, fail to be legitimate probability distributions~\cite{Leonhardt}.
The negativities are thus considered as nonclassicality signatures.
Within multipartite quantum states, the phase-space nonclassicality is tempted to be interpreted as quantum correlations, due to the fact that in a classical description of the joint system no such effects are present~\cite{GlauberBook,VogelBook}.

The sharpest contrast between the definition of quantum correlations in quantum information science and that of quantum optics has been demonstrated very recently by Ferraro and Paris~~\cite{Ferraro12}.
They showed that the two definitions from quantum information and quantum optics are maximally inequivalent, meaning that every quantum state which is classically correlated with respect to the quantum information definition of quantum correlations is necessarily quantum correlated with respect to the quantum optical criteria and vice versa.

One can also compare the operational differences between the two approaches.
On one hand, the quantum correlations of quantum information have been shown to be necessary for specific nonlocal quantum communication~\cite{Dakic12} and computation~\cite{Knill98,Datta08} tasks to outperform their classical counterparts.
On the other hand, however, quantum correlations in quantum optics lack such a nonlocal operational justification, i.e., there is no particular quantum information protocol which exploits phase-space nonclassicality to outperform a classical counterpart protocol.
Although, it has been recently shown that such nonclassicalities provide either necessary or sufficient resources for entanglement generation~\cite{Vogel14,Killoran16,Gholipour16}, which then can be used in various protocols.

In this Letter, we introduce nonlocal \textsc{BosonSampling} as an intermediate model of quantum computing which is performed by distant agents.
We use this specific protocol to fill the gap of an operational interpretation of phase-space nonclassicality in quantum informatics.
Specifically, we show that there exists a quantum state which is strictly classical with respect to entropic measures of correlations in quantum information, allowing for efficient classical simulation of local statistics in our protocol, which at the same time, prohibits efficient classical simulation of nonlocal correlations.
The only known resource present within the state of our example is that of phase-space nonclassicality.
Hence, we see that, nonlocal \textsc{BosonSampling} takes advantage of phase-space nonclassicality to perform a nonlocal task more efficiently than any classical algorithm.


\paragraph*{Correlations in Quantum Information.---}

The main intuition for defining quantum correlations from the viewpoint of quantum information science can be phrased as the following question: 
how much information can be locally encoded into and decoded from a delocalized quantum state?
The answer, upon interpreting entropy as a measure of information, leads to the following unified classification of quantum states based on their correlation type~\cite{Werner89,Openheim02,Horodecki05}:
(i) entangled states; which cannot be prepared locally using any amount of local operations and classical communication (LOCC).
Formally, such states cannot be written in the \textit{separable} form $\hat{\varrho}_{\rm AB}{=}\sum_i p_i \hat{\varrho}_{{\rm A};i}{\otimes}\hat{\varrho}_{{\rm B};i}$.
It has been shown that there are serious restrictions on the amount of information that can be extracted from an entangled state using LOCC~\cite{Badziag03,Horodecki04}.
(ii) Two-way quantum correlated states; that can be prepared using LOCC, however, not all the information can be retrieved using LOCC.
Such states do not allow for any of the local states to be represented via a set of locally distinguishable states (see the next category).
(iii) One-way quantum-classical correlated (QC) states; which can be written as $\hat{\varrho}_{\rm AB}{=}\sum_i p_i \hat{\varrho}_{{\rm A};i}{\otimes}|i\rangle_{\rm B}\langle i|$, or, $\hat{\varrho}_{\rm AB}{=}\sum_i p_i |i\rangle_{\rm A}\langle i|{\otimes}\hat{\varrho}_{{\rm B};i}$, i.e., via a set of locally orthonormal states.
The information encoded within these states cannot be fully recovered via LOCC~\cite{Horodecki05}.
(iv) Strictly classical-classical (CC) states admitting the form $\hat{\varrho}_{\rm AB}{=}\sum_i p_i |i\rangle_{\rm A}\langle i|{\otimes}|i\rangle_{\rm B}\langle i|$.
All the information encoded within a CC quantum quantum state can be decoded using LOCC.
Operationally, a quantum state can be locally broadcasted if and only if it is a CC state~\cite{Piani08}.

The class of CC states simply encode a joint probability distribution $\{p_i\}$ using distinguishable states within the quantum formalism, and thus, are assumed to possess no quantum correlations that gives quantum advantage to a nonlocal information processing task.
The correlation within a quantum state is usually measured using entropic measures, e.g., quantum discord, as follows.
The total amount of correlations within a quantum state is defined to be given by the mutual information as $\mathtt{I}(\hat{\varrho}_{\rm AB}){=}\mathtt{S}(\hat{\varrho}_{\rm A}){+}\mathtt{S}(\hat{\varrho}_{\rm B}){-}\mathtt{S}(\hat{\varrho}_{\rm AB})$, in which $\mathtt{S}(\hat{\varrho}){=}{-}{\rm Tr}\hat{\varrho}{\log}_2\hat{\varrho}$ is the von Neumann entropy and $\hat{\varrho}_{\rm X}$ (${\rm X}{=}{\rm A},{\rm B}$) is the marginal state of $\hat{\varrho}_{\rm AB}$.
Defining the conditional entropy of Bob's subsystem $\mathtt{S}(\hat{\varrho}_{\rm B|A}){=}\mathtt{S}(\hat{\varrho}_{\rm AB}){-}\mathtt{S}(\hat{\varrho}_{\rm A})$, one has $\mathtt{I}(\hat{\varrho}_{\rm AB}){=}\mathtt{S}(\hat{\varrho}_{\rm B}){-}\mathtt{S}(\hat{\varrho}_{\rm B|A})$.
When Alice's subsystem is subjected to a projective measurement $\{\hat{\Pi}^{\rm A}_i\}$, one can vary the conditional entropy of Bob as $\mathtt{S}_{\{\hat{\Pi}^{\rm A}_i\}}(\hat{\varrho}_{\rm B|A}){=}\sum_i p_i \mathtt{S}(\hat{\varrho}_{\rm B|i})$, where $\hat{\varrho}_{{\rm B}|i}$ and $p_i$ are the conditional state and the probability of obtaining the $i$th outcome upon this measurement, respectively.
This gives the new expression for mutual information as $\mathtt{J}_{\{\hat{\Pi}^{\rm A}_i\}}(\hat{\varrho}_{\rm AB}){=}\mathtt{S}(\hat{\varrho}_{\rm B}){-}\sum_i p_i \mathtt{S}(\hat{\varrho}_{\rm B|i})$.
Minimizing our ignorance of Alice's subsystem, we have the mutual information $\mathtt{J}_{\rm A}(\hat{\varrho}_{\rm AB}){=}\sup_{\{\hat{\Pi}^{\rm A}_i\}}\mathtt{J}_{\{\hat{\Pi}^{\rm A}_i\}}(\hat{\varrho}_{\rm B|A})$, which is interpreted as the amount of classical correlations between Alice and Bob.
Quantum discord (from Alice to Bob), is the discrepancy between the two values of total and classical information, $\mathtt{D}_{{\rm A}\to{\rm B}}{=}\mathtt{I}(\hat{\varrho}_{\rm AB}) {-} \mathtt{J}_{\rm A}(\hat{\varrho}_{\rm AB})$~\cite{Henderson01,Ollivier01}.
A quantum state has zero discord from Alice to Bob and vice versa if and only if it is a CC state~\cite{Henderson01,Ollivier01}, i.e., all the information encoded within the state is classical.
Consequently, within the context of quantum information, quantum discord is hypothesised as the most general form of quantum correlations~\cite{Openheim02,Horodecki05}.


\paragraph*{Phase-Space Nonclassicality.---}

Let us now briefly review a second approach to quantum correlations mostly used in quantum optics.
A bosonic system can be described over a quantum analogue of the classical phase-space using the so-called quasiprobability representations.
For this purpose, one defines an informationally complete set of operators, the Weyl-Wigner operators $\hat{T}(\alpha,s)$, as the Fourier transform of the phase-space displacement operator, $\hat{D}(\xi){=}e^{\xi a^\dag - \xi^* a}$,
\begin{equation}
\hat{T}(\alpha,s) = \dmua\xi e^{\alpha\xi^*-\alpha^*\xi + \frac{s}{2}|\xi|^2} \hat{D}(\xi).
\end{equation}
Here, $a$ ($a^\dag$) is the bosonic annihilation (creation) operator, $\alpha,\xi \in \mathbb{C}$ are points of the complex phase-space, and $s\in[-1,1]$ is an arbitrary parameter~\cite{GlauberBook}.
Weyl-Wigner operators are normal, i.e., ${\rm Tr}\hat{T}(\alpha,s){=}1$, and satisfy the orthogonality (duality) relation ${\rm Tr}\hat{T}(\alpha,s)\hat{T}(\beta,-s){=}\pi\delta^{(2)}(\alpha{-}\beta)$.
As a result, we can expand any operator in this basis.
In particular, for $s{=}-1$ one has $\hat{T}(\alpha,-1){=}|\alpha\rangle\langle\alpha|$, with $|\alpha\rangle$ being the bosonic coherent states, and thus, $\hat{\Lambda} {=} \dmua\alpha P_{\hat{\Lambda}}(\alpha) |\alpha\rangle\langle\alpha |$ for an arbitrary operator $\hat{\Lambda}$.
The \emph{real} valued quasiprobability $P_{\hat{\Lambda}}(\alpha)$ is known as the Glauber-Sudarshan $P$-function of the operator $\hat{\Lambda}$.
For values $s{=}0,-1$ one obtains the Wigner and Husimi-Kano quasiprobabilities, respectively.
The generalization of the formalism to bipartite (and multipartite) case is straightforward, and one obtains
\begin{equation}
\hat{\Lambda}_{\rm AB} = \dmub\alpha\beta P_{\hat{\Lambda}_{\rm AB}}(\alpha,\beta) |\alpha\rangle\langle\alpha|\otimes |\beta\rangle\langle\beta|.
\end{equation}

The $P$-function of a quantum state, in general, is not a legitimate probability distribution and can be negative or highly singular.
Such negativities are commonly accepted as nonclassicality signatures of the quantum state~\cite{VogelBook}, and can always be detected using nonclassicality criteria~\cite{Shchukin05,Shchukin05-2,Sperling15}, or can be directly observed via an appropriate smoothing of the $P$-function~\cite{Kiesel08,Kiesel09,Kiesel11}.
However, in the bipartite case, such a nonclassicality interpretation ($P$-nonclassicality) is a source of debate.
The reason is that there exist quantum states for which the marginal $P$-functions are regular probability distributions, while the \emph{joint} $P$-function of the state contains negative regions.
However, these examples range from highly entangled states, such as two-mode squeezed vacuum states, to fully separable ones~\cite{Marek09,Agudelo13}.
Hence, while there are cases in which $P$-nonclassicality can be interpreted to be a signature of quantum correlations, in many other cases there is a question if such an interpretation is well-justified, or if the joint nonclassicality must be considered as a mere consequence of classical correlations?
Importantly, Ferraro and Paris~\cite{Ferraro12} have shown that the two notions of $P$-nonclassicality in quantum optics and quantum discord in quantum information are maximally inequivalent in the sense that a genuine quantum state which is classical according to one notion, is nonclassical according to the other one.
Consequently, if $P$-nonclassicality was proven to be generally a signature of quantum correlations, then, one would have to give up the interpretation of quantum discord as the most general form of quantum correlation, and admit that all non-product bipartite quantum states contain quantum correlations which can be used as a resource in some quantum information task.
At this point, it is important to emphasize that both definitions contain entanglement as a special case: every entangled state of a bosonic system is necessarily $P$-nonclassical and discordant.


\paragraph*{\textsc{BosonSampling} Problem.---}

Intermediate models of quantum computation are those who do not provide universal quantum computation, however, are capable of efficiently solving particular problems which
exhibit an exponential-time speedup compared to classical computers.
\textsc{BosonSampling} is one such a model~\cite{AA}.
Consider a lossless passive linear-optical network (PLON), where $n$ single photons are injected into $m$ input modes of the PLON.
The initial state is given by
\begin{equation}
|\vec{n}\rangle = \bigotimes_{i=1}^{m} (a^\dag_{i})^{n_i} |0\rangle,
\end{equation} 
where $\vec{n}=(n_1,\dots,n_m)$ with $n_i\in\{0,1\}$ characterizes the configuration of the single-photon at the input such that $\sum_{i=1}^m n_i=n$, and $a^\dag_i$ is the bosonic creation operator of the $i$th mode.
As the generator of the unitary transformation corresponding to the PLON is quadratic, its action on the input state, $\hat{U}|\vec{n}\rangle$, can be equivalently described by the symplectic transformation of the mode operators by a matrix $\mathsf{U}$, $\hat{U} a^\dag_j \hat{U}^\dag = \sum_{i=1}^m \mathsf{U}_{ij} a^\dag_i$~\cite{Balian69}.
Assuming that $m\gg n$ (typically, $m\sim O(n^2)$), a particular sample from output distribution can be represented by the vector $\vec{s}=(s_1,\dots,s_m)$ with $s_i\in\{0,1\}$ so that $\sum_{i=1}^m s_i=n$, and occurs with the probability $p(\vec{s}|\hat{U},\vec{n})=|\langle\vec{s}|\hat{U}|\vec{n}\rangle|^2=|{\rm Per}\mathsf{A}(\vec{s};\vec{n})|^2$~\cite{Scheel04}.
Here, ${\rm Per}\mathsf{A}(\vec{s};\vec{n})$ is the permanent of a particular submatrix $\mathsf{A}(\vec{s};\vec{n})$ of the unitary $\mathsf{U}$ determined by the input and sample vectors $\vec{n}$ and $\vec{s}$.

According to Aaronson and Arkipov (AA)~\cite{AA}, generating samples of the output distribution of the \textsc{BosonSampling} device described above is a computationally \#{P}-hard problem, meaning that it cannot be efficiently done on a classical computer. 
Furthermore, AA showed, up to some very feasible conjectures, that even sampling from a distribution close to the exact output distribution of a BosonSampler (in the total variation distance) cannot be efficiently simulated on a classical computer.
Consequently, such a device has a computational power beyond any classical computer, demonstrating the power of (intermediate) quantum computers.


\begin{figure}[h]
  \includegraphics[width=0.8\columnwidth]{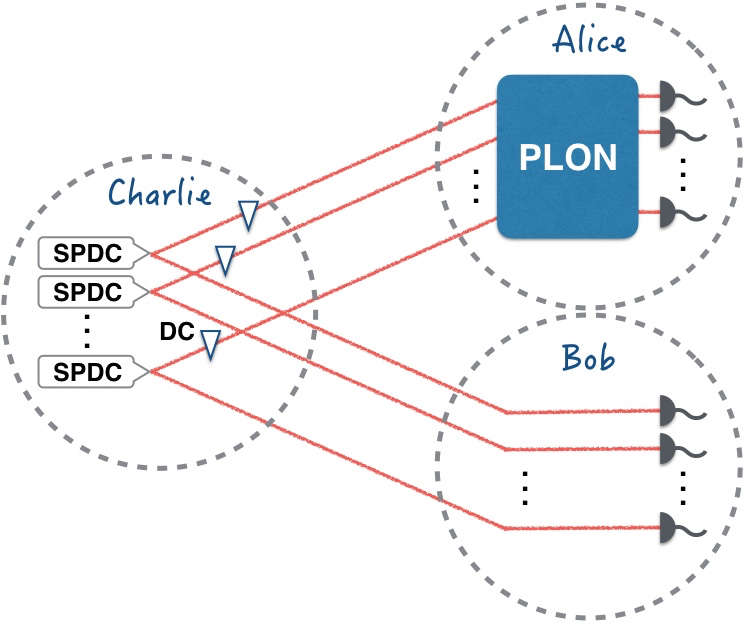}
  \caption{(Color online) The schematic of a nonlocal \textsc{BosonSampling} protocol with CC input state.
  Charlie uses $m$ SPDC sources and a series of dephasing channels (DC) to produce fully dephased two-mode squeezed vacuum states (FDTSV), and shares the final state between two spatially separated agents.
  Alice performs \textsc{BosonSampling} using a passive linear-optical network (PLON) and $\{0,1\}$ Fock basis measurements,
  while Bob only performs $\{0,1\}$ Fock basis measurements.
  We show that, Alice and Bob can simulate their local sample statistics classically efficiently.
  However, they cannot efficiently simulate the correlations between their outcomes using classical computers and an infinite amount of classical communication, although there is no entanglement or discord between agents at any time.
  }\label{Scheme}
\end{figure}

\paragraph*{Quantum Correlations in Nonlocal \textsc{BosonSampling}.---}

We are now ready to present the main result of this Letter.
Consider two agents, Alice and Bob, in two spatially separated laboratories, A and B, respectively.
Alice is equipped with a PLON of size $m$ as depicted in Fig.~\ref{Scheme}.
Moreover, a third party, Charlie, is equipped with $m$ spontaneous parametric down-conversion (SPDC) sources.
The output of the $i$th SPDC is a two-mode squeezed vacuum state, the continuous variable counterpart of a Bell state, and can be written as
\begin{equation}
|\psi_i\rangle = (1-\epsilon_i^2)^{\frac{1}{2}} \sum_{j=0}^{\infty} \epsilon_i^j |j\rangle_{{\rm A};i}|j\rangle_{{\rm B};i},
\end{equation}
where $|j\rangle_{{\rm X};i}$ is the $j$-photon (Fock) state and $0 < \epsilon_i <1$ is the squeezing strength of the $i$th SPDC source.
This state is both entangled and discordant.
To remove these quantum correlations, Charlie applies a dephasing channel with a uniform distribution over all phases $\theta\in[0,2\pi]$ on one output mode of each SPDC.
Mathematically, one has
\begin{equation}\label{FDTSVS}
\begin{split}
\hat{\varrho}_i &= \int_0^{2\pi} d\theta \hat{R}_{\rm A}(\theta)|\psi_i\rangle\langle\psi_i|\hat{R}_{\rm A}(\theta) \\
& = (1-\epsilon_i^2) \sum_{j=0}^{\infty} \epsilon_i^{2j} |j\rangle_{{\rm A};i}\langle j|\otimes|j\rangle_{{\rm B};i}\langle j|,
\end{split}
\end{equation} 
in which $\hat{R}_{{\rm A}_i}(\theta)=e^{-\bold{i}\theta a_i^\dag a_i}$ is the phase shift operator acting on the mode ${\rm A}_i$.
The fully dephased two-mode squeezed vacuum state (FDTSV) of Eq.~\eqref{FDTSVS} is a CC state with no quantum correlations from the quantum information viewpoint, due to the local orthogonality of the Fock states~\cite{Henderson01,Ollivier01}.
The FDTSV state has also other interesting features as shown by Agudelo \textit{et al}~\cite{Agudelo13}:
its marginals, $\hat{\varrho}_{{\rm X};i} = (1-\epsilon_i^2) \sum_{j=0}^{\infty} \epsilon_i^{2j} |j\rangle_{{\rm A};i}\langle j|$ (${\rm X}={\rm A},{\rm B}$), are thermal states with mean photon number $\bar{n}_{\rm th}=\epsilon_i^2/(1-\epsilon_i^2)$, and thus, classical with respect to $P$-function criterion;
nevertheless, its global $P$-function contains negativities, i.e., the global state is $P$-nonclassical.
This is, thus, the only known nonclassical feature of the FDTSV state under consideration.

Next, Charlie sends the first mode of each SPDC source to Alice and the second mode to Bob, as depicted in Fig.~\ref{Scheme}, each of them receiving part of the quantum state $\hat{\varrho}_{\rm AB}=\hat{\varrho}_i^{\otimes m}=(1-\epsilon^2)^m \sum_{j_1,\dots,j_m=0}^\infty \epsilon^{2\sum_{k=1}^m j_k} \left(\bigotimes_{k=1}^m|j_k\rangle_{\rm A}\langle j_k|\right)\otimes \left(\bigotimes_{k=1}^m|j_k\rangle_{\rm B}\langle j_k|\right)$.
Here, without loss of generality, we have assumed that all the squeezing strengths are equal, $\epsilon=\epsilon_1=\dots=\epsilon_m$.
It is now clear that the full quantum state distributed between Alice and Bob, $\hat{\varrho}_{\rm AB}$, is fully separable and fully nondiscortant with classical marginal $P$-functions.
Importantly, the output state after the action of the PLON is also separable and nondiscordant with respect to the Alice-Bob partitioning, and possesses $P$-classical marginals.
This is because the action of the PLON is local and that it maps $P$-classical inputs onto $P$-classical outputs.

Alice inputs the modes she receives into her \textsc{BosonSampling} device, while Bob only makes photon number measurements on the modes he receives.

We now show that Alice and Bob can efficiently simulate their local measurement statistics on their classical computers.
The marginal states that Alice receives for each mode is simply a thermal state possessing a positive $P$-function $P_{{\rm A};i}(\alpha_i)$.
It is well-known that \textsc{BosonSampling} with such input states can be efficiently simulated on a classical computer as follows:
interpret the $P$-function of the inputs, $P_{\rm A}(\alpha_1,\dots,\alpha_m)=\Pi_{i=1}^m P_{{\rm A};i}$, as a joint probability distribution; 
randomly sample from the input phase-space points according to this probability distribution;
apply the corresponding linear transformation of the PLON to the chosen points to obtain the output phase-space points $(\zeta_1,\dots,\zeta_m)$; 
and, calculate the probability of a particular sample $p(\vec{s}_{\rm A}|\hat{U},\alpha_1,\dots,\alpha_m)=e^{-\sum_{i=1}^m |\alpha_i|^2}\Pi_{i=1}^m |\zeta_i|^{2 n_i}$~\cite{Saleh15}.
Bob also receives exactly the same state, except that his PLON is an identity transformation, therefore, he can also efficiently simulate his local measurement statistics.

Can Alice and Bob also simulate their joint measurement statistics efficiently classically, possibly using an infinite amount of classical communication?
This is the problem of nonlocal \textsc{BosonSampling}.
According to our intuition from the classification of quantum correlations in the quantum information context, we conclude that they should be able to simulate coincidence statistics efficiently using a classical computer.
The reason is that, from this viewpoint, there is nothing nonclassical about the correlations stored within the output states:
having access to deterministic photon sources, Alice and Bob could easily encode the joint probability distribution $p_{\rm AB}=\{p_j=\Pi_{i=1}^m\epsilon_i^{2j}\}$ onto their respective states merely using classical communication.
They could also decode all the information stored within the state using local measurements in Fock basis.
This fact is reflected in the non-discordant form of the shared state between Alice and Bob.
The PLONs have also been applied locally, so, there is nothing nonlocal about the operations on the states. 
There is also nothing nonlocal about the measurements; 
they are fully separable in a perfectly distinguishable bases.
On top of that, the events are locally classically efficiently simulable.
Consequently, it seems reasonable not to expect any quantumness of the correlations present in this protocol.
The following theorem, however, proves our naive intuition wrong.

\paragraph*{Theorem.}
The correlations of Alice's and Bob's events cannot be efficiently simulated on a classical computer.

\paragraph*{Proof.} 
It is clear, due to the perfect photon number correlations between pairs of modes shared between Alice and Bob, that Bob's measurements outcomes $\vec{s}_{\rm B}$ heralds a particular configuration of single photons $\vec{n}_{\rm A}=\vec{s}_{\rm B}$ to Alice's BosonSampler with probability $p(\vec{s}_{\rm B}|\hat{\varrho}_{\rm AB})=(1-\epsilon^2)^m\epsilon^{2n}$.
Therefore, Alice's samples statistics \emph{conditioned} on Bob's measurements outcomes is given by $p(\vec{s}_{\rm A}|\hat{\varrho}_{\rm AB},\hat{U},\vec{s}_{\rm B})=p(\vec{s}_{\rm A}|\hat{U},\vec{s}_{\rm B})p(\vec{s}_{\rm B}|\hat{\varrho}_{\rm AB})$.
As shown in Ref.~\cite{Lund14}, this extra randomness from heralding probability does not affect the hardness argument of \textsc{BosonSampling} mapping it onto an equivalently hard problem referred to as scatter-shot \textsc{BosonSampling}, and thus, the conditional statistics $p(\vec{s}_{\rm A}|\hat{\varrho}_{\rm AB},\hat{U},\vec{s}_{\rm B})$ cannot be efficiently simulated on a classical computer.

Similarly, conditioned on a particular sample from Alice's device $\vec{s}_{\rm A}$, one easily finds Bob's probability of a particular measurement outcome to be proportional to the permanent of a submatrix of Alice's PLON matrix, i.e., $p(\vec{s}_{\rm B}|\hat{\varrho}_{\rm AB},\hat{U},\vec{s}_{\rm A})=p(\vec{s}_{\rm A}|\hat{\varrho}_{\rm AB},\hat{U})p(\vec{s}_{\rm B}|\vec{s}_{\rm A})=(1-\epsilon^2)^m\epsilon^{2n}|{\rm Per}\mathsf{A}(\vec{s}_{\rm B}|\vec{s}_{\rm A})|^2 = p(\vec{s}_{\rm A}|\hat{\varrho}_{\rm AB},\hat{U},\vec{s}_{\rm B})$.
Consequently, we can follow the same line of proof as above to conclude that it is impossible to efficiently classically simulate Bob's outcomes statistics \emph{conditioned} on Alice's choice of sample.

Altogether, Alice and Bob cannot simulate the correlations between their events on their classical computers, even by having access to an infinite amount of classical communication.\qed

We emphasize here that, the protocol above and its implications can straightforwardly be generalized to an \emph{approximate} nonlocal \textsc{BosonSampling} scenario in which Alice's PLON corresponds to a Haar random unitary, and the squeezing parameter $\epsilon$ is chosen appropriately~\cite{Lund14}.
Our protocol unveils a new aspect of nonlocal correlations, that is, their computational complexity.
In nonlocal \textsc{BosonSampling}, the quantumness of the correlations is being manifested in the contrast between  the inefficiency of all classical algorithms for simulating the correlations and the efficiency of the nonlocal \textsc{BosonSampling} in a quantum setting.
As noted before, the input and output states in our protocol, as the only sources of any possible correlations, do not meet the quantum-correlations criterion of quantum information, while, at the same time, showing nonclassical correlations.
However, they fulfil the $P$-function nonclassicality criterion~\cite{Agudelo13}.
This observation leads us to the conclusion that negativity of $P$-function provides the resource for nonlocal \textsc{BosonSampling} and can be interpreted as a signature of true quantum correlations beyond entanglement and discord.


\paragraph*{Conclusion.---}

We have explored the quantumness of correlations within quantum states beyond entanglement and quantum discord, by considering the setting of a nonlocal \textsc{BosonSampling}.
We showed that there exists a bipartite quantum state containing no entanglement or discord, and considered to be strictly classical within the context of quantum information, yet makes our nonlocal \textsc{BosonSampling} efficient.
At the same time, we showed that sampling from the local output distributions of our device with our particular input and output states can be efficiently classically simulated, while the joint output distribution cannot.
In short, we concluded that there exists quantum correlations within our input and output states that are used as a resource to outperform any classical algorithm in simulating the correlations between outcome events.
We then noticed that our state does satisfy the quantum-correlation criterion of quantum optics, i.e., it has a $P$-nonclassical phase-space representation, and thus such nonclassicalities can be considered as the resource in our protocol.


\acknowledgements{The authors gratefully acknowledge Andrew G. White who stimulated this research.
This project was supported by the Australian Research Council Centre of Excellence for Quantum Computation and Communication Technology (CE110001027).}

\end{document}